\documentstyle[prl,aps,amsfonts,amssymb,twocolumn,epsfig]{revtex}

    \renewcommand{\vec}[1]{{\bf #1}}
    
    \newcommand{\ccfa}{CeCu$_5$Au }

    \newcommand{\crg}{CeRu$_{2}$Ge$_2$ }

    \newcommand{\tn}{$T_{\rm N}$ }

\begin{document}

\title{Pressure--induced residual resistivity anomaly in CeCu$_5$Au}

\author{H. Wilhelm, S. Raymond, and D. Jaccard}
\address{DPMC, Universit\'e de Gen\`eve, Quai E.-Ansermet 24, CH-1211 Geneva 4,
  Switzerland}

\author{O. Stockert 
\footnote{Present Address: H. H. Wills Physical Laboratory, 
University of Bristol, Tyndall
 Avenue, Bristol, BS8 1TL, Great Britain} and H. v. L\"ohneysen}
\address{Physikalisches Institut, Universit\"at Karlsruhe, D-76128 Karlsruhe, 
Germany }

\author{A. Rosch}
\address{Institut f\"ur Theorie der Kondensierten Materie, Universit\"at 
Karlsruhe, D-76128 Karlsruhe, Germany}

\date{\today}

\wideabs{
\maketitle


%
%
\begin{abstract}
  The electrical resistivity of the magnetically ordered \ccfa has been
  investigated under pressure up to 8.5\,GPa. In the magnetically ordered
  region ($p<3.4$\,GPa) the residual resistivity $\rho_0$ shows a pronounced
  maximum as a function of pressure. Even in the nonmagnetic region $\rho_0$
  decreases monotonically by more than a factor of three. These two effects
  can be qualitatively explained in terms of the interplay of pressure,
  magnetism and disorder in a strongly correlated electron system with
  weak disorder.
\end{abstract}

\pacs{72.15.-v, 75.20.Hr, 72.10.Di, 62.50.+p}


} 

%
%
The behavior of the electrical resistivity $\rho$ of metals when approaching
zero temperature $T$ has been of interest ever since the beginning of solid
state physics. While it is clear that the residual resistivity $\rho_0$ of
conventional metals is largely governed by lattice defects and impurities, the
situation is much less clear for metals with strong electronic correlations
and/or magnetic order at low $T$.  In the independent-electron approximation,
$\rho_0 = m^*/e^2n \tau = \frac{3}{2} \pi (h/e^2)/(k_{\rm }F^2 l)$, where
$m^*$ is the effective mass, $\tau$ the scattering time arising from electron
scattering by defects and impurities, $l$ the corresponding mean free path,
$n$ the electron density and $k_{\rm F}$ the Fermi wave-number.  In
heavy-fermion (HF) systems, $m^*$ is enhanced by up to a factor of 1000 due to
the Kondo effect arising from the exchange coupling of conduction electrons
and nearly localized $f$-electrons. Nevertheless, $\rho_0$ is rather low in
stoichiometric compounds: the quasi particles obey Bloch's theorem and scatter
at $T=0$ only from rare defects.  The simple independent-electron expression
above suggests that hydrostatic pressure $p$ should affect $\rho_0$ only to a
minor extent since both $k_{\rm F}$ and $l$ should depend on $p$ only through
the (small) change of electron density and interatomic distances.

In this letter, we report $\rho_0$ measurements under hydrostatic pressure on
the stoichiometric HF compound CeCu$_5$Au where heavy quasiparticles, leading
to a Sommerfeld coefficient $\gamma = 0.64$\,J/molK$^2$ for $T\rightarrow 0$,
coexist with incommensurate antiferromagnetic (AF) order \cite{PASCH94}. The
magnetic ordering temperature $T_{\rm N}$ in this system can be suppressed by
pressure \cite{WILHE00}. Surprisingly, a  strong dependence of $\rho_0$ on
$p$ is found: starting from $\rho_0 = 28\,\mu\Omega {\rm cm}$ for $p$ = 0,
$\rho_0$ passes over a pronounced maximum of 58\,$\mu\Omega {\rm cm}$ at
1.8\,GPa and decreases to 20\,$\mu\Omega {\rm cm}$ at $\approx$ 3\,GPa.
Even in the nonmagnetic Fermi-liquid state ($p\gtrsim
3.4$\,GPa), $\rho_0$ decreases substantially upon further pressure increase,
reaching 6\,$\mu\Omega {\rm cm}$ at 8.5\,GPa.

%
%

The single crystal of \ccfa (space group {\it Pnma}) used in this study was
grown with the Czochralski method from the starting constituents Ce (5N), Cu
(5N), and Au (4N). A part of the crystal was used for neutron experiments
which revealed a very good quality of the single-phase sample \cite{STOCK99}.
Two small rectangular pieces ($35\times 73\times 861\mu$m$^3$ and $28\times
85\times 561 \mu$m$^3$) cut from the same crystal were used in two pressure
experiments. A clamped-anvil high-pressure device \cite{JACCA98}, capable of
reaching 10~GPa, was used to measure $\rho(T)$ by the four-point method for a
current $I$ along the crystallographic $b$ direction, i.e. perpendicular to
the magnetic ordering vector $\vec{Q}$ \cite{STOCK99}.  The sample and the
pressure manometer (a thin Pb foil) were embedded in a soft
pressure-transmitting medium (steatite) to ensure homogeneous pressure
conditions. Both samples showed the same $\rho_0(p)$ behavior, ruling out an
error due to a possible rearrangement of the contact leads under pressure.

%
%

The $\rho(T)$ data of the larger sample are shown in Fig.\,1. The AF phase
transition at ambient $p$ occurs at $T_{\rm N}=2.35$~K, in perfect agreement
with specific-heat and magnetization measurements \cite{PASCH94}.  Reducing
the unit--cell volume by external pressure tunes \tn to zero \cite{WILHE00}.
This behavior is well known for a number of HF systems and is attributed to an
increase of hybridization between conduction electrons and local moments, thus
favoring the Kondo effect over the RKKY interaction. The detailed analysis of
these data \cite{WILHE00} reveals a critical pressure $p_c \approx$ 3.4\,GPa
for the quantum-critical point (QCP) where $T_{\rm N} \rightarrow 0$.

Figure 2 shows the pressure dependence of $\rho_0$, determined in the $T$
range 30~mK $<T<50$~mK. The particularly striking observations are (i) the
strong nonmonotonic variation of $\rho_0$ within a few GPa where $T_{\rm N}$
is monotonically suppressed to zero, and (ii) the strong decrease 
of $\rho_0$ even in the paramagnetic state (for $p>3.4$\,GPa). 
\begin{figure}
\begin{center}
\epsfxsize=80mm \epsfbox{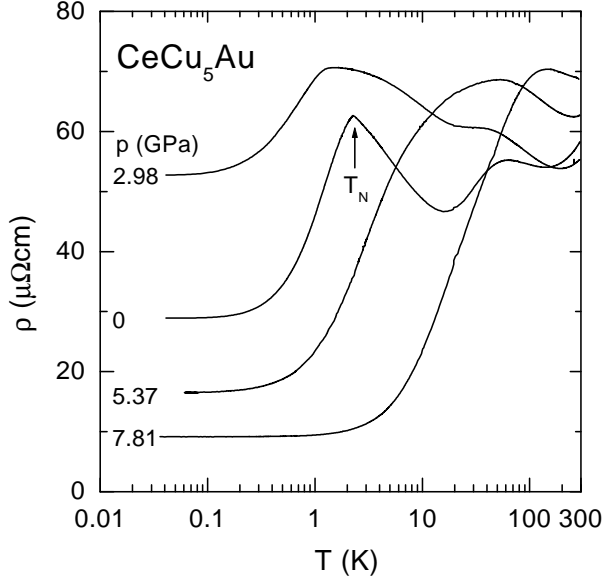}
\end{center}
\caption{Temperature dependence of the electrical resistivity  $\rho$ of CeCu$_5$Au
at selected pressures $p$ in a semilogarithmic scale.}
\end{figure}
\noindent
The maximum in $\rho_0(p)$ is centered at $p \approx 1.8$~GPa, i.~e., well
inside the magnetic phase.  The $\rho_0(p)$ in the magnetic region is in a
superficial regard qualitatively similar to the $\rho_0(x)$ variation in
CeCu$_{6-x}$Au$_x$ \cite{LOEHN98}. However, $\rho_0(x)$ with a maximum for $x
= 0.5$ arises from $x$-dependent disorder and negative lattice pressure while
external pressure is not expected to change the number of impurities in
CeCu$_5$Au.

Pressure is not the only parameter to change $\rho_0$ drastically. Likewise,
an external magnetic field $B$, applied along the (easy) crystallographic $c$
direction, yields a reduction of $\rho$ at low $T$. Figure\,3 shows $\rho(T)$
curves at $p$ = 2.2 and 3.5\,GPa for $B$ = 0 and 7.5~T. Well inside the
magnetic region (for $p$ = 2.2~GPa, $T_{\rm N}=1.35$~K), $B$ = 7.5~T destroys
the AF order and causes a decrease of $\rho_0$ by more than 60\%.  In the
nonmagnetic phase ($p$ = 3.5~GPa), $B$ affects $\rho(T)$ as well, but here the
influence becomes weaker as $T \rightarrow 0$.  Hence, the ``magnetic''
contribution to $\rho_0$ decreases as the system is tuned away from the
magnetic-nonmagnetic instability.

The impressive $\rho_0(p)$ dependence of \ccfa should be regarded in the
context of the $p$-induced $\rho_0$ changes in other HF systems. For
CeCu$_2$Ge$_2$ a maximum is found well inside the nonmagnetic phase as for
CeCu$_2$Si$_2$ \cite{JACCA99}. The magnitude of the $\rho_0$ peak in
superconducting CeCu$_2$Ge$_2$ at a pressure close to the $T_{\rm c}(p)$
maximum, can be associated with charge fluctuations \cite{MIYAK99} or even a
valence transition. For YbCu$_2$Si$_2$ only a fraction of the Yb ions orders
magnetically at $p_{\rm c}$ \cite{WINKE99} and the broad and shallow
$\rho_0(p)$ maximum may correspond to full ordering at higher $p$. In the case
of \crg the $\rho_0(p)$ variation at low $p$ might be related to a change of
the magnetic ordering vector between low and intermediate $p$ \cite{WILHE99},
but
\begin{figure}
\begin{center}
 \epsfxsize=80mm \epsfbox{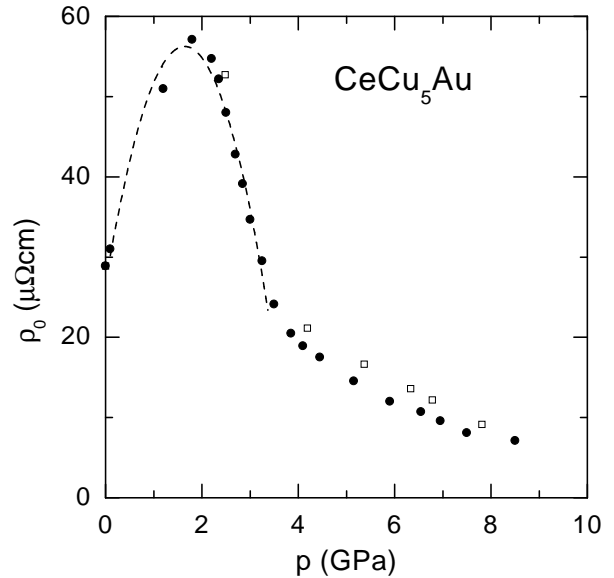}
\end{center}
\caption{Residual resistivity $\rho_0$ of \ccfa vs. pressure $p$
  for a current $I$ along the $b$ axis. The two symbols denote 
results obtained on different samples. The line is a guide to the eye.}
\end{figure}
\noindent
the effect is rather small.

%
%
%
%

The following discussion treats our major observations for CeCu$_5$Au, i.e.,
the (i) nonmonotonic $\rho_0(p)$ dependence in the AF phase and (ii) the
strong $\rho_0$ decrease for $p>p_c$, in terms of models that are generic to
strongly correlated systems with some disorder.

We first focus on the AF phase and argue that the strong $\rho_0(p)$
dependence is a consequence of the quasi-particle scattering from local
variations in the staggered magnetization.  The qualitative picture is as
follows: impurities change the chemical environment of the Ce atoms and lead
to a large spatial variation of the local susceptibilities $\chi_{\rm
  L}(\vec{r})\approx 1/T_{\rm K}(\vec{r})$ with a ``Kondo temperature''
$T_{\rm K}(\vec{r})$ which depends {\em exponentially} on the local
environment. In the AF phase this leads to a variation of the local
magnetization $\Delta M_{\rm L}(\vec{r})\propto f[H_{\rm AF}/T_{\rm
  K}^{\text{imp}}]-f[H_{\rm AF}/T_{\rm K}^{\text{bulk}}]$, where $H_{\rm
  AF}(\vec{r})$ is a staggered (exchange) field proportional to the AF order
parameter and $f[x]$ describes the crossover from weak magnetization $f[x\to
0]\propto x$ to a saturation of the magnetic moment for large $ H_{\rm AF}$
(upper inset of Fig.\,4).  The electrons scatter off the variations of the
(local) magnetization $\Delta M_{\rm L}$.  Close to the QCP, $|\Delta M_{\rm
  L}|$ increases with increasing staggered bulk magnetization $M_{\rm AF}$.
Deep in the AF phase, the local magnetization saturates, $H_{\rm AF}\gtrsim
\max[T_{\rm K}^{\text{imp}},T_{\rm K}^{\text{bulk}}]$, and $|\Delta M_{\rm
  L}|$ decreases in size. This explains qualitatively the pronounced maximum
of $\rho_0(p)$. It also corresponds to the observation that close to the
$\rho_0(p)$ maximum, $T_{\rm N}$ is of the order of the coherence temperature,
measured by the maximum in $\rho(T)$ \cite{WILHE00}. This signals a smooth
crossover from a spin-density wave with small magnetic moments 
just below $p_c$ to local-moment AF order with large mo-
\begin{figure}
\begin{center}
  \epsfxsize=80mm \epsfbox{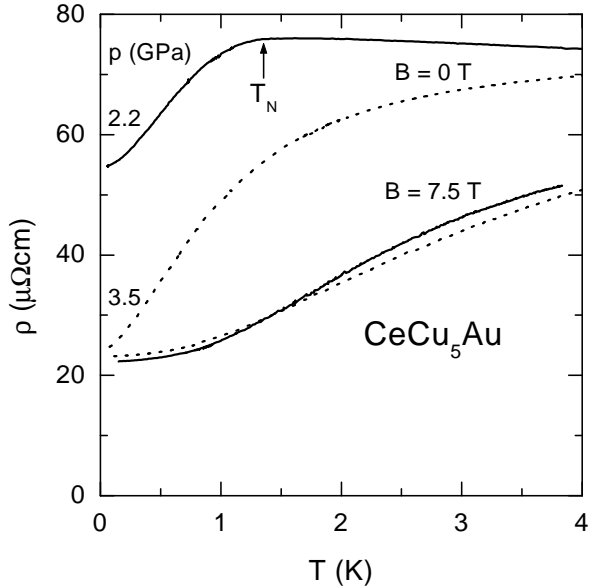}
\end{center}
 \caption{Electrical resistivity $\rho(T)$ of \ccfa for $p=2.2$~GPa (solid lines)
   and $p=3.5$~GPa (dotted lines) in zero magnetic field and $B =7.5$\,T.}
\end{figure}
\noindent
ments as $p$ approaches zero.

To substantiate this qualitative picture we have performed a microscopic
calculation \cite{ROSCH00} for an Anderson-lattice model in the limit $U\to
\infty$: $H=\sum_{ij, \sigma} t_{ij} c^\dagger_{i\sigma} c_{j\sigma} + V
c^\dagger_{i\sigma} f_{i\sigma}+ \epsilon_f^i f^\dagger_i f_i + U
f^\dagger_{i\uparrow}f_{i\uparrow} f^\dagger_{i\downarrow}f_{i\downarrow}$. We
mimic changes of $p$ by a variation of the hybridization $V$. Increasing the
values of $V$ corresponds to a $p$ increase. The local physics of this
strongly correlated system was treated in the dynamical mean-field theory
(MFT), the resulting self-consistent Anderson model was solved within
slave-boson MFT \cite{GEORGES96}. While this approximation fails to describe
the inelastic processes correctly, it is expected to give a qualitatively
correct description for low $T$. The disorder is modeled by a small density of
$f$-electron sites with a different local energy
$\epsilon_{f}^i=\epsilon_f^{\text{imp}}\neq \epsilon_f^{\text{bulk}}$.  For
simplicity, the long-range order is described by a staggered magnetic field
with ordering wave vector $\vec{Q}$ = (1/2 0 0). In Fig.\,4, $\Delta\rho_0$ in
a direction perpendicular to $\vec{Q}$ (like in the experiment) as a function
of $M_{\rm AF}$ shows indeed the maximum which we expected from the arguments
given above.  The fact that we obtain a sharp cusp instead of a smooth maximum
in $\Delta\rho_0(M_{\rm AF})$ is a well known artefact of the MFT treatment of
the single-impurity Anderson model where a spurious phase transition is
induced when the magnetization is close to saturation.

The approximation used above neglects non-local effects which should be
important in the AF phase close to the QCP, when the correlation length $\xi$
is large.  We estimate this effect in perturbation theory by considering the
scattering from non-local variations of the staggered magnetization $\Delta
M(\vec{r}) \approx \int \chi(\vec{r}-\vec{r}') \Delta M_{\rm
  L}(\vec{r}')/\chi_{\rm L} d \vec{r'}$, where $\chi$ is the susceptibility of
the AF phase, characterized by $\xi$.  
\begin{figure}
\begin{center}
\epsfxsize=80mm  \epsfbox{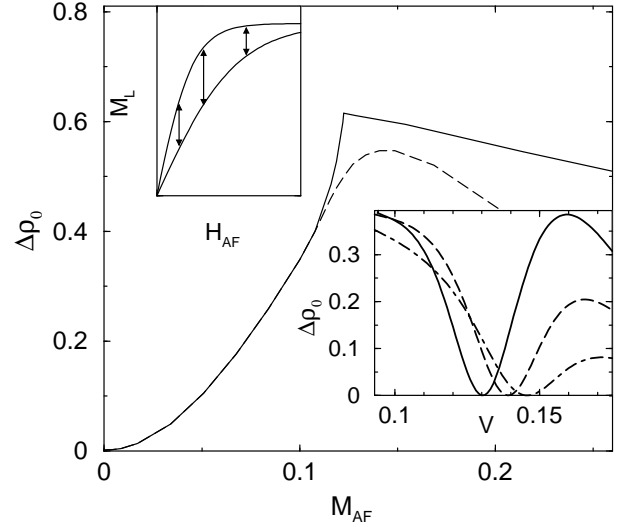}
\end{center}
\caption{Upper inset: Schematic plot of the local magnetization $M_{\rm L}$
  as a function of a local magnetic field $H_{\rm AF}$ at a bulk site and an
  impurity site.  In the AF phase the electrons scatter from local variations
  of the magnetization $\Delta M_{\rm L}= M_{\rm L}^{\text{imp}}-M_{\rm
    L}^{\text{bulk}}$ which leads to the pronounced maximum in $\Delta\rho_0$,
  shown in the main figure as a function of the bulk magnetization $M_{\rm
    AF}$ (for $\epsilon_f^{\text{imp}}< \epsilon_f^{\text{bulk}}<0$). The cusp
  in the solid line is an artefact of the approximation used. The dashed line
  shows schematically the smooth maximum expected in a more realistic theory.
  Lower inset: $\Delta\rho_0$ in the paramagnetic phase as a function of the
  hybridization $V$ (larger $V$ corresponds to higher $p$) for three
  different values of $\epsilon_f^{\text{imp}}>\epsilon_f^{\text{bulk}}$. }
\end{figure}
\noindent
For independent impurities, small
$\Delta M$ and $\Delta \rho_0 \ll \rho_0$ \cite{ROSCH99a}, we find an
enhancement of $\Delta \rho_0$ for large $\xi$, $\Delta \rho_0 \propto \xi
|\Delta M_{\rm L}|^2$.  Unfortunately, there are no data on the $p$ dependence
of the magnetization $M_{\rm AF}$ or $\xi$. From MFT, we expect $\Delta M_{\rm
  L} \propto \sqrt{p_c-p}$ and $\xi \propto 1/\sqrt{p_c-p}$ and therefore
$\Delta \rho_0 \sim \sqrt{p_c-p}$ close to $p_c$. This is not observed as
$\rho_0(p)$ appears to rise linearly with $p_c-p$ and may be attributed to the
fact that MFT strongly overestimates the increase of the size of $M_{\rm AF}$
as suggested by the $x$-dependence of $M_{\rm AF}$ in CeCu$_{6-x}$Au$_x$
\cite{LOEHN98}. Also the assumption of independent impurities breaks down
close to the QCP.  Furthermore, $\xi$ may be too short in the relevant $p$
regime (inelastic neutron scattering at the QCP of CeCu$_{6-x}$Au$_x$ reveals
relatively broad structures \cite{STOCK98}).
 
Other potentially important effects like the opening of a gap in some parts of
the Fermi surface and an associated reduction of scattering rates
\cite{LOEHN98} have been included in our model calculation. We do not expect
them to dominate the $p$ dependence of $\rho_0$ measured {\em perpendicular}
to $\vec{Q}$. Likewise, our qualitative picture should hold also for the
incommensurate ordering vector $\vec{Q}$ = (0.56 0 0) for CeCu$_5$Au
\cite{LOEHN98}. A semi-quantitative check for the local picture is a
comparison with the magnetoresistivity at the QCP, where according to our
local approximation a similar rise of $\rho_0$ is expected for not too strong
fields. Indeed, in CeCu$_{5.9}$Au$_{0.1}$, $\rho_0$ (measured for $I$ along
the $b$ direction) rises by approximately $20\%$ for a local magnetization of
$0.3 \mu_{\rm B}$ \cite{NEUBERT}. This is of the expected order of magnitude. A more
precise comparison to our pressure data is not possible due to the different
nature of the impurities in CeCu$_{5.9}$Au$_{0.1}$. Finally, we note that the
strong field dependence in the range where $\rho_0(p)$ is large, is in
qualitative agreement with our model.

%
%
%
%
%

We now turn to the pronounced $\rho_0(p)$ dependence in the paramagnetic
region of the phase diagram which is indeed surprising from a theoretical
point of view since in a Kondo lattice with nonmagnetic impurities, $\rho_0$
is only weakly renormalized by interactions \cite{KOTLIAR90}.  It is tempting
to interpret the increase of $\rho_0$ upon approaching the QCP from the
paramagnetic side as the signature of the spontaneous formation of magnetic
domains of size $\xi$ around certain disorder configurations. Such an
interpretation seems not viable as the same strong variation of $\rho_0$ is
observed at a considerable distance from the QCP, where the number of those
domains should be exponentially small.  We have therefore theoretically
investigated the question whether a strong $p$ dependence may arise from
purely local effects.

Each impurity, i.e., a Kondo ion with a local environment differing from the
bulk due to disorder, contributes according to Friedel's sum rule a factor
$\Delta \rho \sim \sin^2 \pi \Delta n$ to $\rho_0$ if $s$-wave scattering
dominates and the density of impurities is small.  $\Delta n$ is the total
charge accumulated (in the $s$-wave channel) around the impurity. For the
following argument we consider an impurity site where the $f$-electron is more
strongly coupled to the environment than a typical bulk site
($\epsilon_f^{\text{bulk}}<\epsilon_f^{\text{imp}}<0$).  For high $p$ or large
$V$, the occupation of the $f$ orbital at the impurity site is therefore
reduced and $\Delta n$ is negative. For small $V$ both the impurity site and
bulk sites are deep in the Kondo regime and the number of $f$-electrons per
site and spin is fixed at $1/2$.  Nevertheless, extra charge is accumulated
around sites with the larger local $T_{\rm K}$. For a {\em single-impurity}
Kondo model this effect is of order $T_{\rm K}/E_{\rm F}$, with $E_{\rm F}$
the Fermi-energy, and therefore negligible. For a lattice $\Delta n$ is of
order $1$ and positive for a particle-like Fermi surface.  In a generic
situation one therefore expects a sign change in $\Delta n$ (for a
particle-like Fermi surface) and correspondingly a {\em strong} $p$ dependence
of $\Delta \rho_0 \sim \sin^2 \pi \Delta n$ as is shown in the lower inset of
Fig.\,4. A similar argument holds in the case
$\epsilon_f^{\text{bulk}}>\epsilon_f^{\text{imp}}$.  Remarkably, in the limit
$T_{\rm K}^{\text{imp}}\ll T_{\rm K}^{\text{bulk}}$ (or $T_{\rm
  K}^{\text{imp}}\gg T_{\rm K}^{\text{bulk}}$) $\Delta n$ takes on a universal
value which depends only on bulk properties \cite{ROSCH00}. 

With this mechanism, it is possible to explain the unexpected large variation
of $\rho_0(p)$ in the paramagnetic phase as an intrinsic property of a weakly
disordered Kondo lattice. We obtain changes of $\rho_0$ by a factor of $\sim
3$ accompanied by changes of $V$ by a few percent and variations of $T_{\rm
  K}$ by a factor of 2-3, in reasonably good agreement with experiment. Note
that our analysis does not contradict the results of Kotliar and Varma
\cite{KOTLIAR90} who find that in a Kondo lattice $\rho_0$ is only weakly
renormalized by interactions due to nonmagnetic impurities: the strong
$\rho_0(p)$ variations are indeed ``weak'' as they are not of order $1/T_{\rm
  K}$.  A prediction of this scenario is that $\rho_0$ rises again for even
higher $p$ upon entering the mixed-valence regime.  From the fact that
$\rho_0(p)$ in CeCu$_6$ decreases for $p$ up to 8.5\,GPa \cite{RAYMO00}, we
deduce that in CeCu$_5$Au such an increase, if present, might occur at
pressures well above $10$\,GPa.

In conclusion, we have observed a surprisingly strong variation of the
residual resistivity with pressure in both the antiferromagnetic and
paramagnetic phase of CeCu$_5$Au. We argued that the pronounced $\rho_0(p)$
maximum in the AF phase can be attributed to scattering from inhomogeneities
in the magnetization. The explanation for the strong decrease of $\rho_0$ in
the paramagnetic phase by a factor of three is less obvious: a possible mechanism
can nevertheless be found even in a purely local picture where the sign of the
charge, which is accumulated in a $s$-wave channel, changes under typical
conditions

%
%

This work was supported by the Swiss National Science Foundation and by the
Deutsche Forschungsgemeinschaft through grant Lo 250/20-1 and the Emmy Noether
program.

%
%
\vspace*{5mm}
\small{
$\ast$ Present Address: H. H. Wills Physical Laboratory, 
University of Bristol, Tyndall
 Avenue, Bristol, BS8 1TL, Great Britain.}

\end{document}